\documentstyle[prb,aps,multicol,epsf]{revtex}
\input epsf

\begin{document}
\title{
Coulomb Correlations  and Pseudo-gap Effects in a Pre-formed 
Pair Model for the Cuprates }
\author{Jiri Maly, K. Levin and D. Z. Liu}
\address{
James Franck Institute, University of Chicago, Chicago, IL 60637}
\address{\rm (Submitted to Physical Review B August 27, 1996)}
\maketitle

\makeatletter
\global\@specialpagefalse
\def\@oddhead{REV\TeX{} 3.0\hfill Levin Group Preprint, 1996}
\let\@evenhead\@oddhead
\makeatother

\begin{abstract}
We extend previous work on pre-formed pair models of 
superconductivity to incorporate Coulomb correlation effects.  For 
neutral systems, these models have provided a useful scheme which 
interpolates between BCS and Bose Einstein condensation with 
increasing coupling  and thereby  describes some aspects of 
pseudo-gap
phenomena.  However, charge fluctuations (via  the plasmon, 
$\omega_p$)  significantly modify the collective modes and
therefore the interpolation behavior. We discuss the resulting 
behavior
of the pseudo-gap and thermodynamic quantities 
such as $T_c$, $\chi$ and $C_v$ as a function of $\omega_p$.
\end{abstract}

\vspace{-0.2in}
\begin{multicols}{2}
The role of the pseudo-gap\cite{phystoday} in the high $T_c$ cuprates is 
emerging 
as an important indicator of the nature of the superconductivity as 
well as the normal state. 
There are two widely discussed but competing explanations for 
pseudo-gap effects but no clear and decisive 
experiments to support  one scenario over the other.
Early observations  associated the pseudo-gap with magnetic 
pairing\cite{magnetic}  
above $T_c$ (often called the ``spin gap''). 
It is now clear, however,
that some form of normal state pairing  is seen in photoemission as 
well as charge transport data.  Moreover, at least in the photoemission 
data the pseudo-gap appears to have the $d$-wave symmetry 
\cite{phystoday} of the 
ordered state and this leads naturally to the association of this ``gap''  
with precursor superconductivity.\cite{nozieres,randeria,emery}
This second scenario is further supported by the observation of low 
dimensionality and short coherence lengths in high $T_c$  
superconductors, 
which suggests important deviations from ideal mean field or BCS 
transitions.  Indeed, the approach of the present paper assumes
the precursor superconductivity scenario,
in large part because it is important to establish, 
at least as a base-line,  the extent to which such superconducting
``fluctuation''  effects may be responsible for pseudo-gap behavior. 

Among those models which subscribe to a precursor 
superconductivity scenario
there are additionally two rather distinct viewpoints.  
Emery and Kivelson\cite{emery} have argued that the pseudo-gap 
state of the 
cuprates is similar to that observed in 
granular films where phase coherence is not fully established, 
although large regions of the material have a well established 
superconducting  amplitude.  Because it is small, in some sense,
in the cuprates their approach focuses on $n/m^*$ or 
alternatively on the plasma frequency $\omega_p$ as the key 
``phase stiffness'' parameter.  Alternatively, others 
\cite{nozieres,randeria,haussmann,simulations} have 
focused on the observed small size of the superconducting 
correlation length $\xi$ to argue for important corrections to BCS 
theory associated with pre-formed or nearly-formed pairs\cite{uemura} which 
exist well above $T_c$ and therefore give rise to significant 
pseudo-gap effects.
The present paper is based on the viewpoint that in the cuprates
the characteristic parameter of the charge degrees of 
freedom, $n/m^*$ or equivalently $\omega_p$, 
should be treated on a relatively equal 
footing  with  the correlation length, $\xi$.

To study the role of Coulomb interactions on pseudo-gap phenomena,
we adopt a natural microscopic framework  
which incoporates charge fluctuations into 
theories which treat the cross-over from BCS pairing to Bose-Einstein 
condensation (BEC) of pre-formed pairs.  In neutral systems, this 
cross-over
has been studied 
by a variety of investigators.\cite{nozieres,randeria,haussmann}
Numerical simulation studies,\cite{simulations} which 
have been performed
in the context of attractive Hubbard models, include, in principle,
all diagrammatic
contributions. On the other hand, analytical work has mostly been 
confined to the T-matrix approximation. The issue of non-conserving and 
conserving T-matrix schemes has 
been widely discussed in the literature\cite{haussmann} in the 
context of the
BCS-BEC cross-over problem. In the original work of 
Nozi\`{e}res and Schmitt-Rink a non-conserving approach was used. 
Recent work\cite{haussmann} on neutral systems has extended this 
scheme using a T-matrix 
approximation which satisfies  global conservation laws and in the 
process introduces renormalized Green's functions into the  
generalized susceptibilities.  
In the charged system, as a consequence 
of gauge invariance,  the analogue  renormalized susceptibilities must 
then appear in the particle-hole channel. As  has been 
known for some time,\cite{kadanoff} however, the collective mode 
spectrum
is then treated incorrectly at this
level of approximation and a more sophisticated scheme is needed.
In order to avoid this complexity and to develop an intuitive 
understanding of the effects of
charge, however, we restrict the analysis, in this paper, to the more 
familiar
scheme introduced by Nozi\`{e}res and Schmitt-Rink and defer 
consideration
of a fully conserving formalism. We note, however, that our formulation 
will be locally 
conserving and in the case of charged systems
this approximation does not yield qualitatively 
different physics from that expected using a globally conserving 
approach.  Furthermore, it is our contention that mode-mode 
coupling 
effects, which are ignored in all T-matrix based schemes,  will 
ultimately lead to important insights which we will discuss in  a 
future paper. 

The Hamiltonian under consideration contains an attractive 
interaction $V_{{\bf k},{\bf k}'}$, parameterized by a coupling 
constant 
$g$, as 
well as long range Coulomb terms.  For definiteness we take the same 
separable pairing potential, $V_{{\bf k},{\bf k}'} = 
gv_{\bf k}v_{{\bf k}'}$
where $v_{\bf k} = (1+k^2/k^2_0)^{-1/2}$, as was used 
initially by Nozi\`{e}res and Schmitt-Rink.  
Within this model, the pairing energy scale (or ``Debye frequency'') is 
the 
Fermi energy. 
We assume a three dimensional free electron  model for the 
electrons and defer discussion of anisotropy effects until later 
in the text. It is assumed  that in the cuprates there is sufficient 
inter-layer hopping so that a strictly two dimensional model and its 
associated Kosterlitz-Thouless transition is not the 
appropriate starting point. 
We generalize the path integral
formulation of Ref. \onlinecite{randeria}, replacing the usual 
fermionic fields by Nambu spinors,
$\psi^{\dagger}_{\bf k}=(c^{\dagger}_{{\bf k}\uparrow},
c_{-{\bf k}\downarrow})$, and decoupling in four real fields
given by $\eta_i = \psi^{\dagger}\tau_i\psi$ ($\tau_i, i=0\ldots3,$ 
are the identity and
three Pauli matrices). The interaction in the off-diagonal channels 
($i=1,2$)
is  the pairing interaction while the Coulomb 
term appears in the diagonal channels ($i=0,3$). Finally, 
the field $\eta_0$ may
be eliminated by a suitable gauge transformation and
the thermodynamic potential 
$\Omega$ is computed for the remaining degrees of freedom 
at the Gaussian approximation level,
\begin{eqnarray*}
\Omega & = &  \Omega_0 + \frac{T}{2}\sum_{{\bf q},\nu_m}\ln\,\det
\left(\Gamma_{{\bf q},\nu_m}^{-1}\right) 
- \frac{n_f}{2}\sum_{\bf q}V^{(3)}_{\bf q}, \\
\Gamma^{(i,j)\,-1}_{{\bf q},\nu} & = & \delta_{i,j} - T\,
V^{(i)}_{\bf q}\sum_{{\bf k},\omega_l}
{\rm tr}(\tau_i\,{\cal G}_{{\bf k+q}/2,\omega_l}\,
\tau_j\,{\cal G}_{{\bf k-q}/2,\omega_l-\nu_m})\,v^{(i)}_{\bf 
k}v^{(j)}_{\bf k}.
\end{eqnarray*}
Here  $\Omega_0$ is the usual contribution from non-interacting 
fermions
and $n_f$ denotes the corresponding number density of free 
fermions. 
In the interaction $V^{(i)}_{\bf q}=g,\ 
v^{(i)}_{\bf k} = v_{\bf k}$ for $i=1,2$,
and $V^{(3)}_{\bf q}=4\pi e^2/\kappa q^2,\ v^{(3)}_{\bf k}=1$.
Finally, ${\cal G}_{{\bf q},\omega}$ is the single particle Green's 
function 
in Nambu space and $T$ is the temperature, while $\nu_m$
and $\omega_l$ are the even and odd Matsubara frequencies, 
respectively.
Above $T_c$ the fluctuation propagator $\Gamma$ 
is diagonal; thus all three channels 
(the particle-particle, particle-hole 
and uncorrelated fermions)  contribute additively to
$\Omega$ as well as to the various derived thermodynamic 
quantities
which we calculate below.  

In the present formalism,\cite{randeria}  the 
transition temperature is obtained from the BCS gap equation with a 
self consistently determined chemical potential, $\mu$, obtained  
from
the condition  $n_{tot} = -\partial\Omega/\partial\mu$.
The resulting coupled equations are solved numerically for $\mu$ 
and $T_c$.
In the limit of arbitrarily large $g$ the effects of the Coulomb 
interaction drop out, since mode-mode coupling is neglected 
at the Gaussian level, so that  $T_c$ is given by the ideal BEC 
temperature. 
At small $g$, the collective mode 
contribution to $\mu$ becomes arbitrarily small and the BCS limit is 
approached, albeit with a Coulomb renormalized chemical 
potential.\cite{mahan}
When the superconducting coupling constant 
$g$ vanishes, the above  form for $\Omega $ reflects the plasmon 
contribution and 
reduces to that of the usual Coulomb gas.\cite{mahan} 
The Gaussian approximation to $\Omega$ gives an
RPA-like treatment of the collective modes and so
provides a reasonable 
scheme for interpolating between these two limits. 
It should be noted, however, that 
Coulomb pseudo-potential effects (which would act to renormalize 
$g$) are, for simplicity, not included in our calculations. Here we 
focus principally  on the effects introduced by  the long range 
Coulomb interaction (which enters via the parameter
${\omega_p}^2 = 4 \pi {n e^2} / (m^*\kappa)$).

\begin{figure}
\vspace{-0.1in}
\narrowtext
\epsfxsize=2.5in 
\hspace{0.5in}\epsfbox{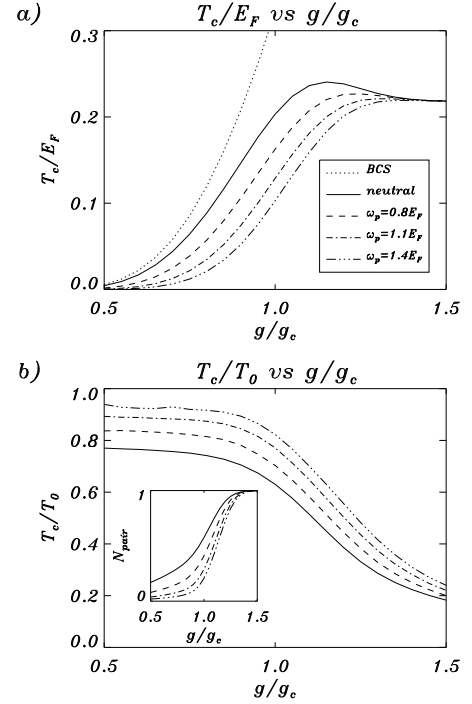}
\caption{
(a) The variation of $T_c$ as a function of coupling constant in the 
BCS  and Gaussian approximation theories for neutral and 
charged systems with increasing plasma 
frequencies. (b) $T_c$ normalized by the Coulomb mean field result
(see text) $T_0$ for the same plasma 
frequencies as in (a). The inset 
in (b) 
plots the number of paired electrons vs $g/g_c $ for the same 
parameter set.
\label{fig-1}}
\end{figure}
To illustrate the effects of charge fluctuations we plot, in 
Fig.~\ref{fig-1}a, $T_c$ as a function of $g/g_c$  
for various values of the plasma frequency. The dotted 
line represents the BCS result (for the neutral system) and the solid 
line is the corresponding (neutral) transition temperature which 
includes Gaussian fluctuations.  The remaining curves from left to 
right demonstrate the effects of increasingly large $\omega_p$. 
For the purposes of focussing 
on charging effects only  we fix $E_F$ and $k_F/k_0$;
in this way all the Gaussian 
derived curves have the same high $g$ asymptote. Moreover, with 
these 
assumptions the neutral system reference curve is unchanged as
$\omega_p$ is varied. Two effects of 
Coulomb interactions can be observed in Fig.~\ref{fig-1}a: (i) $T_c$ 
decreases with increasing $\omega_p$,  and (ii)  the non-monotonic
behavior as a function of $g$ found for the 
neutral case 
(which is believed to be unphysical)\cite{haussmann} progressively 
disappears with increasing $\omega_p$.  

The first observation, which is perhaps the more important,
is a consequence of the fact that an attraction
due to Coulomb interactions in the particle-hole channel 
reduces the effectiveness of the attraction in the pairing
channel.  We illustrate  this first point more directly 
in the inset of Fig.~\ref{fig-1}b, which plots  the number of superconducting
pairs as a function of increasing $\omega_p$. 
This effect may seem counter to the expectation that systems with 
larger $\omega_p$ will have reduced $T_c$ suppression (i.e. larger $T_c$)
due to phase 
fluctuations.  However, when the appropriate reference temperature 
is used, 
Coulomb interactions  are found to lead to better agreement with 
mean field 
theory; in this sense superconducting  fluctuations are, indeed, 
suppressed
by Coulomb interactions. We plot in the main portion of Fig.~\ref{fig-1}b 
the ratio of  $T_c$ 
to the critical temperature $T_0$ obtained by neglecting pair 
fluctuations (but including Coulomb effects)
for the same range
of plasma frequencies as in Fig.~\ref{fig-1}a.  
As can be seen, the larger the plasma frequency, the 
higher the ratio $T_c / T_0$  and thus the better the agreement  
with a mean field treatment of the pairing channel. 

\begin{figure}
\vspace{-0.1in}
\narrowtext
\epsfxsize=2.5in 
\hspace{0.5in}\epsfbox{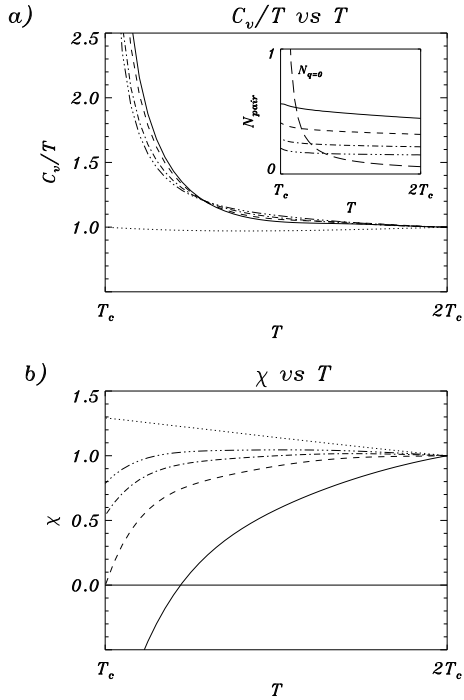}
\caption{
Temperature dependent specific heat (a) and spin susceptibility 
(b) for the same parameters as in Figures 1. The dotted lines are the 
BCS results (with slightly temperature dependent chemical potential).  
The 
inset in (a) plots the total number of pairs vs $T$, along with the 
number at $q=0$.
\label{fig-2}}
\end{figure}
A more convenient way of illustrating  fluctuation effects and 
the associated role of the plasma frequency,  however,  is to study 
thermodynamic properties directly. Here pseudo-gap effects enter  
as precursor superconducting contributions in, for 
example, the specific heat and spin susceptibility. As in Fig.~\ref{fig-1}b, 
these fluctuation effects are expected to weaken as the 
plasma frequency increases and mean field behavior is restored.  
We plot $C_v$ and $\chi$ in Fig.~\ref{fig-2} for the same parameter set as in 
Fig.~\ref{fig-1}, with $g/g_c$ set equal to unity. This choice of coupling 
strength
is consistent with the observation of relatively short coherence 
lengths 
in the cuprates and with the claim that high $T_c$ compounds lie 
close 
to the bound state 
limit.\cite{uemura}  Here the dotted lines again represent the BCS 
value (with a properly
$T$ dependent $\mu$).  As is appropriate\cite{thouless} for  
Gaussian 
fluctuation theories, the specific heat varies as $(T-T_c)^{-1/2}$. 
Comparison with the neutral (solid) reference curve 
shows that the effects of variable $\omega_p$ are not as evident in 
the specific heat as in the spin susceptibility.  Similarly, deviations
from mean-field behaviour appear at higher temperatures  in $\chi$ 
than 
in $C_v$.

It is worth noting that in the
present formulation there is no well defined ``onset temperature''
for the appearence of a pseudo-gap, as might exist
if a sharp phase transition occured at some temperature $T^*$ above 
$T_c$.
Moreover in discussing the onset of fluctuations, it should be stressed 
that  their appearance  is unrelated to the temperature dependence of 
the total number of fluctuations or pre-formed
pairs.  As shown in the inset of Fig.~\ref{fig-2}a,  for fixed $\omega_p$, 
the total number of pairs is relatively constant up to
temperatures many times higher than $T_c$; however, their distribution
shifts to lower momenta as $T_c$ is approached and long range 
coherence
is established. Detailed calculations indicate that varying
$\omega_p$ does
not alter the critical behaviour in the pairing channel as
$T_c$ is approached; the narrowing of the pseudo-gap region 
with increasing $\omega_p$ is a result of the smaller relative contribution
that pairing fluctuations make to the thermodynamics as Coulomb
correlations become more dominant. Finally, it should also be noted 
that the neutral system 
yields unphysical thermodynamic behavior at smaller 
coupling constants than when Coulomb effects are included. 
This is illustrated  by 
the negative values of $\chi$ indicated in the figure. 
Such unphysical behaviour has been
shown to result from the non-conserving nature of the 
approximations used.\cite{sofo}

While the above figures were designed to illustrate the effect of 
Coulomb correlations, they do not fully represent the physical system 
in which variations in the plasma frequency are necessarily 
associated with changes in the electronic energy scale. In reality both 
$\omega_p$ and $E_F = {k_F} ^2 / 2m^* $ 
depend on similar combinations of the carrier density $n$ and 
effective mass $m^*$.  As the insulator is approached $\omega_p$ 
decreases as $ {\omega_p } ^ 2 \approx x $ (where $x$ denotes the 
number of doped holes); however,  whether one 
assumes a Fermi liquid ($n\approx 1+x$) or non-Fermi
liquid  ($n \approx x$) approach to the insulator, it follows 
necessarily that  the 
electronic energy scale $E_F$ must also vanish as the hole 
concentration $x$ approaches $0$.
In scenarios 
based on electronic pairing mechanisms, therefore, it is difficult 
to escape the 
conclusion  that the onset temperature for coherent superconducting
fluctuations, $T^*$, 
should  also  become small as the insulator is approached. Our 
numerical calculations of $C_v$ and $\chi$ exhibit this effect, 
principally because ours is a single energy scale theory: 
both $T^*$ and $T_c$ derive from the same pairing mechanism. This 
behavior  is in contrast to experiment\cite{phystoday} where  even 
in highly underdoped systems  $T^*$ is of the order of $100K$ or 
more.   Large $T^*$ seems to be most naturally associated with a high 
energy scale in the  insulating parent 
compound, such as a magnetic energy.\cite{magnetic} However, in such 
a scenario it then becomes problematic to understand how the other 
important energy scale\cite{uemura} $\omega_p$ enters to determine 
$T_c$. We speculate that a cross-over from three to two dimensionality 
may play some role in  lowering $T_c$ at low doping concentrations 
where enhanced  (quasi-2d) critical fluctuation 
effects are most apparent.\cite{loram}

\begin{figure}
\vspace{-0.1in}
\narrowtext
\epsfxsize=2.5in 
\hspace{0.35in}\epsfbox{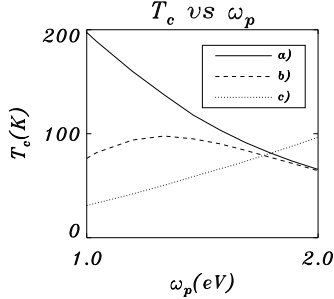}
\caption{
Variation of $T_c$ with $\omega_p$ (in units of $eV$, where optimal doping
corresponds to $\omega_p\approx 1.2eV$) with
$E_F$ held fixed (curves (a) and (b)) as well as using 
self-consistently determined  $E_F$ (curve(c)). Curve (b) corresponds
to $2+\epsilon$ dimensions with $\epsilon=0$ when $\omega_p=0$
and $\epsilon=1$ near optimal. See text
for details.
\label{fig-3}
}
\end{figure}
In Fig.~\ref{fig-3},
we explore these  issues  within the context of our model. 
We consider three situations in which the Gaussian derived $T_c$ is 
plotted as a function of $\omega_p$ (which may be directly related to 
hole concentration $x$) to arrive at a form of ``phase diagram''.  
The solid  line corresponds to the case in which the characteristic 
energy for pairing (called ``$E_F$'') is held fixed, and as observed 
in Fig.~\ref{fig-1}a, $T_c$ monotonically decreases.  
This should be contrasted 
with the situation in which  $E_F$ is allowed to vary
self-consistently (dotted line) in accord with the measured $\omega_p$. Here
for definiteness, we assume a Luttinger volume Fermi surface. 
The latter case yields\cite{uemura} $T_c \approx \omega_p$;  however, 
$T_c$ is suppressed at low doping primarily as a result 
of the lowering of the electronic energy scale,  $E_F$, rather than 
from increased phase fluctuations. Neither of the above cases is 
entirely satisfactory for understanding the larger pseudo-gap regime 
at low doping.  We address this issue phenomenologically by introducing  
the effects of a dimensionality crossover for fixed pairing energy scale  
$E_F$ and
calculating  $T_c$ for a system in $2+\epsilon$ dimensions
(dashed curve) where $\epsilon$ varies smoothly from zero
when $\omega_p\approx 0$ (at half filling)  to one when 
$\omega_p\approx 1.2eV$ 
(corresponding to optimal doping).  Here $T_c$ is 
suppressed at low doping, as a consequence of 2d fluctuation effects. 
Moreover, one can associate  $T^*$ with the solid curve,  which exhibits the 
observed experimental trends.  Thus the low doping regime  is characterized 
by a large pseudo-gap. On the other hand, at higher doping or $\omega_p$,
both  $T_c$
and $T^*$ converge and  the pseudo-gap region vanishes. Whether or not 
this phenomenology is appropriate, the above discussion underlines the 
importance of multiple energy scales ($\omega_p$, as distinct from the pairing
energy scale)  and the possible role of a dimensionality crossover\cite{loram}
in understanding pseudo-gap behavior.

In summary, we have presented a pre-formed pair  model in which 
the 
effects of Coulomb correlations  are clearly seen to suppress 
superconducting fluctuations (above $T_c$) and thereby tune 
pseudo-gap behavior in the calculated specific heat and spin 
susceptibility. Moreover, these two thermodynamic variables, as a 
function of 
$T$, are found to be reasonably consistent with experiment. 
However, 
it should be stressed that within our microscopic model, 
the effects of Coulomb correlations enter in a rather different way
than has been assumed in previous phenomenological 
schemes.\cite{emery,doniach}  Our approach focusses more 
directly on correlated pairs rather than superconducting grains;  
therefore, phase and amplitude fluctuations appear on a 
relatively equivalent basis.  
As a consequence, introducing Coulomb interactions into the 
pre-formed pair formalism
leads to a narrowing of the pseudo-gap region  
by providing a competing attraction in the particle-hole channel
and thus reducing the effectiveness of superconducting pairing. 

This work is 
supported by the National Science Foundation (DMR 91-20000) 
through the
Science and Technology Center for Superconductivity.
KL acknowledges the  hospitality and support,  via the 
National Science Foundation (PHY94-07194),  of the Santa Barbara 
Institute for Theoretical Physics and JM greatfully acknowledges the
financial support of the National Science and Engineering Research
Council (Canada).  Useful conversations with C. Castellani, 
M. Randeria, C. Sa de Melo and M. Tarlie are acknowledged.

\end{multicols}
\end{document}